\documentclass[twocolumn,aps,pra,showpacs,showkeys,superscriptaddress,floatfix]{revtex4-1}
\usepackage[T1]{fontenc}
\usepackage[utf8]{inputenc}
\usepackage{amsmath}
\usepackage{amssymb}
\usepackage{esint}
\usepackage{graphicx}
\usepackage{color}
\usepackage{subfigure}

\makeatother

\begin{document}

\title{Entanglement of Two Jaynes-Cummings Atoms In Single Excitation Space }

\author{Ya Yang}
\affiliation{Key Laboratory of Low-Dimensional Quantum Structures and Quantum Control of Ministry of Education, Department of Physics and Synergetic Innovation Center of Quantum Effects and Applications, Key Laboratory for Matter Microstructure and Function of Hunan Province, Hunan Normal University, Changsha
410081, China}

\author{Yan Liu}
\affiliation{College of Physics and Electronic Engineering, Hengyang Normal University, Hengyang 421002, China}

\author{Jing Lu}
\thanks{Corresponding author}
\email{lujing@hunnu.edu.cn}
\affiliation{Key Laboratory of Low-Dimensional Quantum Structures and Quantum Control of Ministry of Education, Department of Physics and Synergetic Innovation Center of Quantum Effects and Applications, Key Laboratory for Matter Microstructure and Function of Hunan Province, Hunan Normal University, Changsha
410081, China}

\author{Lan Zhou}
\affiliation{Key Laboratory of Low-Dimensional Quantum Structures and Quantum Control of Ministry of Education, Department of Physics and Synergetic Innovation Center of Quantum Effects and Applications, Key Laboratory for Matter Microstructure and Function of Hunan Province, Hunan Normal University, Changsha
410081, China}

\begin{abstract}
We study the entanglement dynamics of two atoms coupled to their own Jaynes-Cummings cavities in single-excitation space. Here we use the concurrence to measure the atomic entanglement. And the partial Bell states as initial states are considered. Our analysis suggests that there exist collapses and recovers in the entanglement dynamics. The physical mechanism behind the entanglement dynamics is the periodical information and energy exchange between atoms and light fields. For the initial Partial Bell states, only if the ratio of two atom-cavity coupling strengths is a rational number, the evolutionary periodicity of the atomic entanglement can be found. And whether there is time translation between two kinds of initial partial Bell state cases depends on the odd-even number of the coupling strength ratio.

\end{abstract}


\date{\today}

\maketitle

\section{Introduction}

It is well known that entanglement is a typical quantum property of compound systems. It plays an essential role in quantum information science, such as quantum computation, quantum cryptography, quantum communication, and quantum measurement \cite{Horodecki2009}. However, quantum entanglement is very fragile, since the entangled systems are unvoidable to interact with their surrounding environments \cite{Breuer2002}. The decoherence is recognized as a main obstacle to realizing quantum information processing \cite{Suter2016}.

In recent years, the dynamical behavior of entanglement under the action of the environment has obtained extensive research \cite{Walter13,Ludwig10,Joshi12,Togan10,Chathavalappil19}. Yu and Eberly have shown that two initially entangled and afterward not interacting qubits can become completely disentangled in a finite time \cite{Yu04,Yu06,Yu09}. This phenomenon is usually called “entanglement sudden death (ESD)”, and has been detected in the laboratory \cite{Kimble2007}. Subsequently, the creation or rebirth of entanglement in a two-qubit system has been found \cite{Ficek06,Ficek08}. Later, the dynamical properties of entanglement for three-qubit states has also been investigated \cite{Ou07,Qiang18}.

The Jay-Cummings (JC) model describes the coherent interaction between a two-level atom and a single radiation mode \cite{Jaynes1963}. In the single excited subspace, the JC model is equivalent to a two-qubit system. As one of few exactly solvable models, the JC model has been exploited for the study of entanglement dynamics. The purpose of this paper is to study the entanglement properties of a system consisting of two isolated two-level atoms in their own JC cavity. These two atoms does not interact but are entangled with each other. Each two-level atom is in a perfect single-mode resonator, but each is completely isolated from the other atom and the other cavity. It is found that the entanglement dynamics of the two atoms is related to the initial entanglement magnitude between two atoms and the atom-cavity coupling strengths. Besides, the sudden death and rebirth of entanglement can also appear under some initial conditions.

The structure of the paper is organized as follows. In Section~II, we introduce the physical model and derive the basic equations for the entanglement dynamics. In Section~III, we investigate in detail the time evolution of the quantum entanglement of two JC atoms for the case of initial partial Bell states. Finally, we conclude this work in Section~V.

\section{\label{Sec:2}The Model and Basic Equations }

\begin{figure}
  \centering
  \includegraphics[width=6cm]{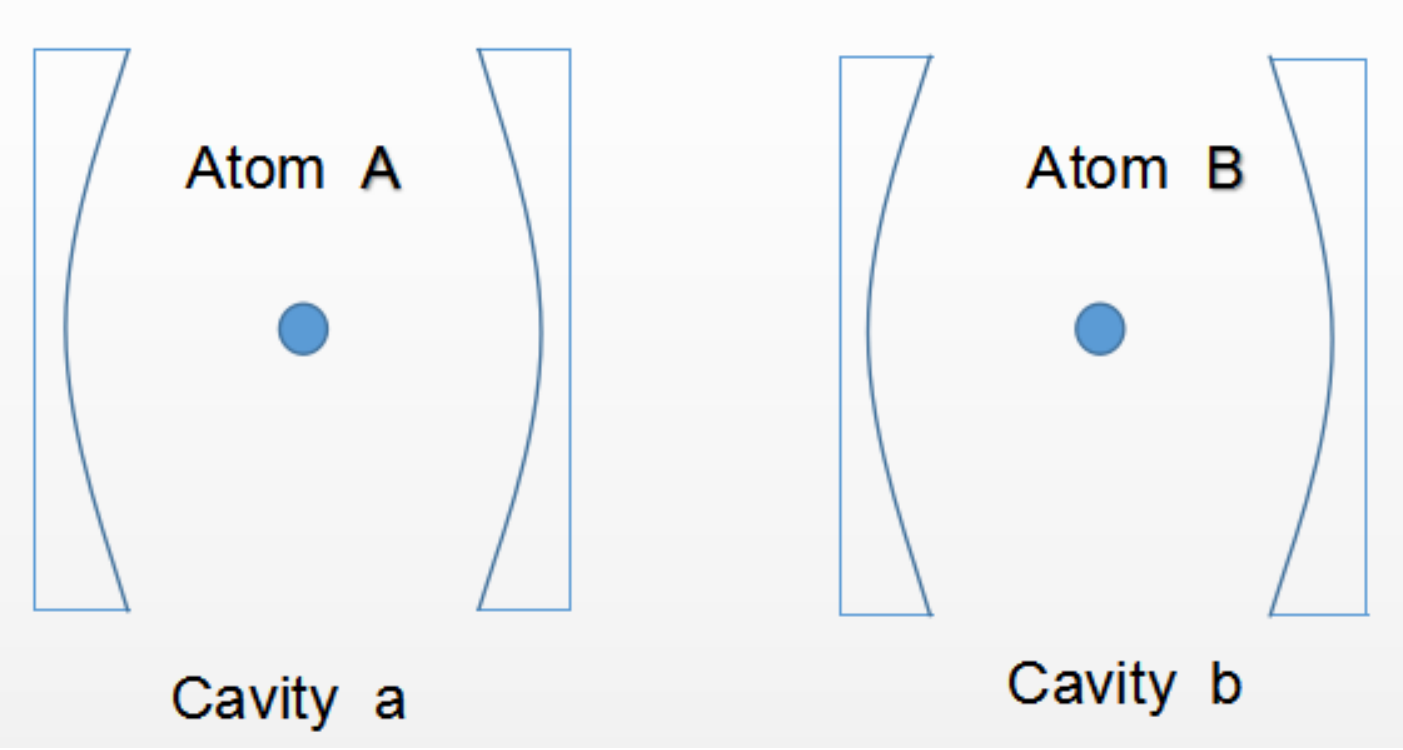}
  \caption{The “double Jaynes-Cummings” model consists of two atoms in their own perfect single-mode resonator cavities. These two atoms does not interact but are entangled with each other. Each two-level atom is completely isolated from the other atom and the other cavity.}\label{fig1}.
\end{figure}

In this section, we consider the system consisting of the double JC model, as schematically shown in Fig.~\ref{fig1}. The Hamiltonian of the system can be described by~\cite{Jaynes1963,Eberly1980,Yonac2006}$ (\hbar=1)$%
\begin{eqnarray}
\label{eq-1}
H &=&\frac{1}{2}\omega \sigma _{A}^{z}+\frac{1}{2}\omega \sigma
_{B}^{z}+\omega _{0}a^{\dagger }a+\omega _{0}b^{\dagger }b  \nonumber \\
&&+g_{A}\left( a\sigma _{A}^{+}+a^{\dagger }\sigma _{A}^{-}\right)
+g_{B}\left( b\sigma _{B}^{+}+b^{\dagger }\sigma _{B}^{-}\right) .
\end{eqnarray}
Here $\omega _{0}$ is the frequency of single-mode cavity a and b, $\omega$ is the transition frequency of two-level atom A and B. $g_{A}$ ($g_{B}$) is the coupling strength between the atom A (B) and optical cavity a (b). $\sigma _{\alpha}^{z}$, $\sigma _{\alpha}^{+}$, $\sigma _{\alpha}^{-}$ are respectively the atomic Pauli z-operator, raising operator and lowering operator for atom $\alpha=A,B$. $a$ ($b$) and $a^{\dagger }$ ($b^{\dagger }$) are the annihilation and creation operators for cavity a (b).

Because the atoms only interact with their own cavities, the eigenstates of this total Hamiltonian are products of the dressed eigenstates of separate JC systems~\cite{Jaynes1963,Eberly1980,Yonac2006}. Note that the total excitation number $N=N_{A}+N_{B}$ is conserved with $N_{A}=a^{\dag}a+\sigma_{A}^{z}$ and $N_{B}=b^{\dag}b+\sigma_{B}^{z}$ being the excitation number of the first and second JC model. Now we just consider that the total excitation number is one with $N=1$, there exist only two categories of eigenstates. The first one is that the excitation exists in the first JC system and the other JC system is in the ground state with $N_{A}=1, N_{B}=0$. The second type is the excitation in the second JC systems with $N_{A}=0, N_{B}=1$. Under the resonance condition with $\omega=\omega_{0}$, the four eigenstates and eigenvalues in the interaction picture are as follows~\cite{Boca04,Yonac07}
\begin{eqnarray}
\label{eq-2}
\lambda _{1}^{\pm }&=&\pm g_{A}, \ \ \left\vert \Psi _{1}^{\pm }\right\rangle = \frac{1}{\sqrt{2}}\left( \left\vert \uparrow _{A}0_{a}\right\rangle |\pm
\left\vert \downarrow _{A}1_{a}\right\rangle \right) \left\vert \downarrow
_{B}0_{b}\right\rangle;\nonumber\\
\lambda _{2}^{\pm } &=& \pm g_{B},\ \  \left\vert \Psi _{2}^{\pm }\right\rangle =\frac{1}{\sqrt{2}}\left( \left\vert \uparrow _{B}0_{b}\right\rangle \pm
\left\vert \downarrow _{B}1_{b}\right\rangle \right) \left\vert \downarrow
_{A}0_{a}\right\rangle.
\end{eqnarray}
In the following, the states are abbreviated as $\left\vert ABab\right\rangle$ with $A,B=\uparrow$ or $\downarrow$, and $a,b=0$ or $1$. Then the bare basis in the single excitation subspace can be rewritten as $\{\left\vert \uparrow \downarrow
00\right\rangle ,\left\vert \downarrow \uparrow 00\right\rangle ,\left\vert
\downarrow \downarrow 10\right\rangle ,\left\vert \downarrow \downarrow
01\right\rangle \}$.
In the subspace, the state at any time reads
\begin{equation}
\label{eq-4}
\left\vert \Psi \left( t\right) \right\rangle =x\left\vert \uparrow
\downarrow 00\right\rangle +y\left\vert \downarrow \uparrow 00\right\rangle
+z\left\vert \downarrow \downarrow 10\right\rangle +k\left\vert \downarrow
\downarrow 01\right\rangle.
\end{equation}
with initial condition $\{x_{0},y_{0},z_{0},k_{0}\}$. Inserting Eqs.~(\ref{eq-1}) and (\ref{eq-4}) into Schr$\rm{\ddot{o}}$dinger equation, the time derivative of the coefficient can be obtained as
\begin{eqnarray}
i\dot{x}=g_{A}z,\ \ \  i\dot{z}=g_{A}x,
\nonumber \\
i\dot{y}=g_{B}k,\ \ \  i\dot{k}=g_{B}y.
\end{eqnarray}
We note that $x$ and $z$ form a closed equation system, and the same is true for $y$ and $k$. This is because there is no interaction between two JC models. Thus, the coefficients can be derived as the following time-dependent formulas,
\begin{eqnarray}
x &=&x_{0}\cos \left( g_{A}t\right) -iz_{0}\sin \left( g_{A}t\right) ,
\nonumber  \label{eq-5} \\
y &=&y_{0}\cos \left( g_{B}t\right) -ik_{0}\sin \left( g_{B}t\right) ,
\nonumber \\
z &=&z_{0}\cos \left( g_{A}t\right) -ix_{0}\sin \left( g_{A}t\right) ,
\nonumber \\
k &=&k_{0}\cos \left( g_{B}t\right) -iy_{0}\sin \left( g_{B}t\right) .
\end{eqnarray}
From Eqs.~(\ref{eq-2})-(\ref{eq-5}), we can see that there must be only one independent JC model evolves over time, while the other is in the ground state $\left\vert \downarrow 0\right\rangle$.

\begin{figure*}[htbp]
\centering
\subfigure[]{\includegraphics[width=7cm]{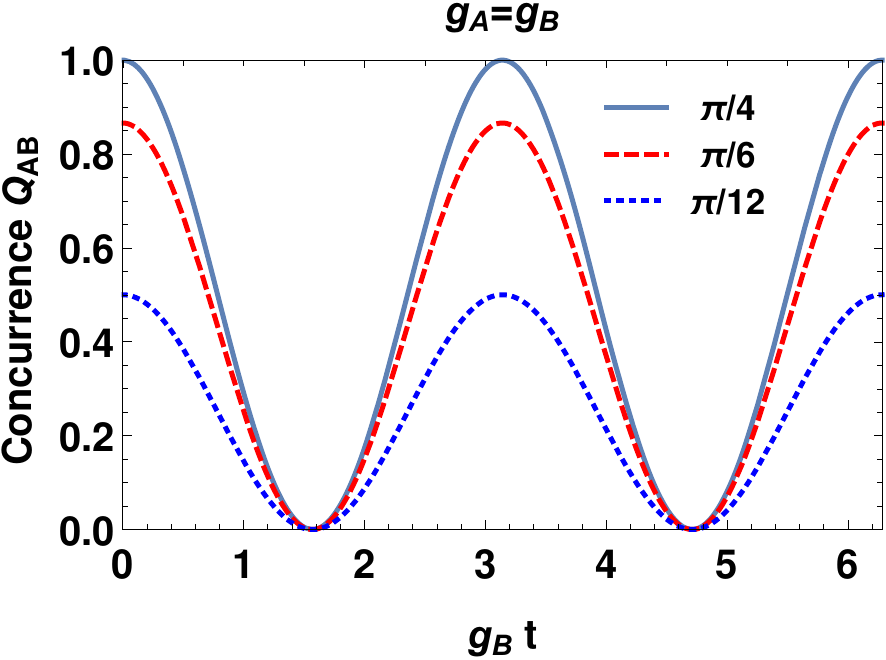}}
\subfigure[]{\includegraphics[width=7cm]{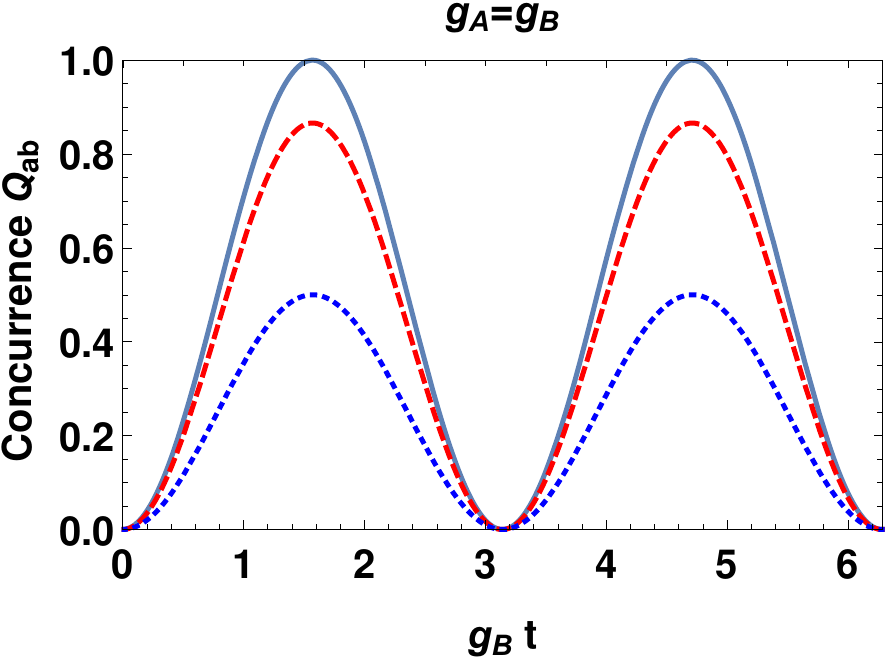}}
\subfigure[]{\includegraphics[width=7cm]{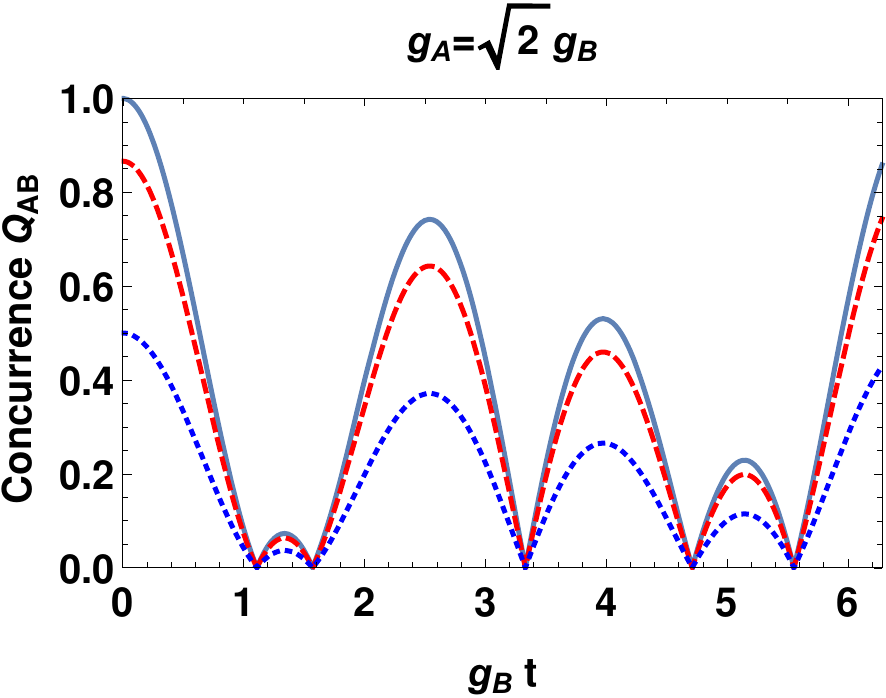}}
\subfigure[]{\includegraphics[width=7cm]{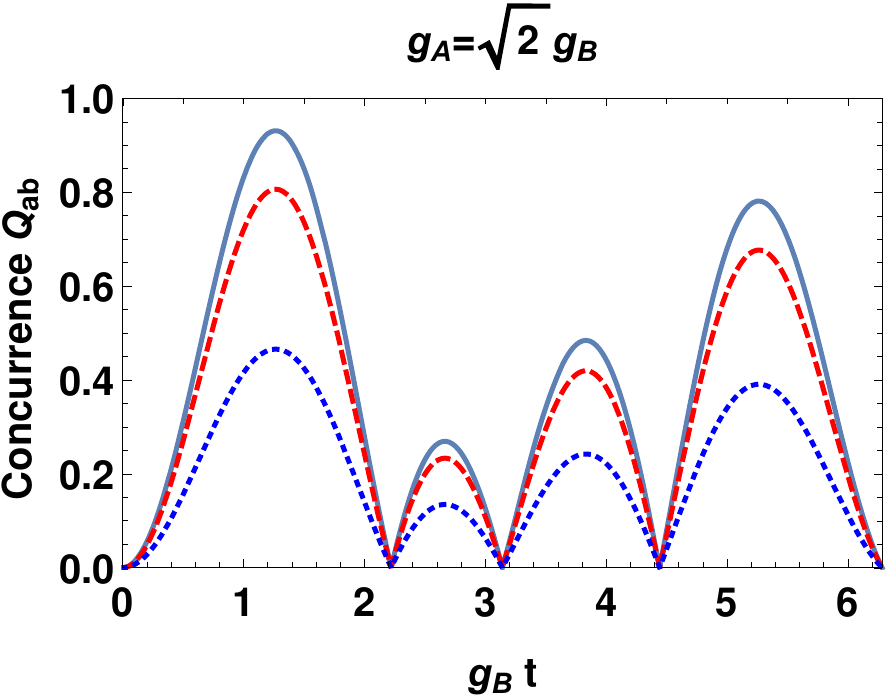}}
\subfigure[]{\includegraphics[width=7cm]{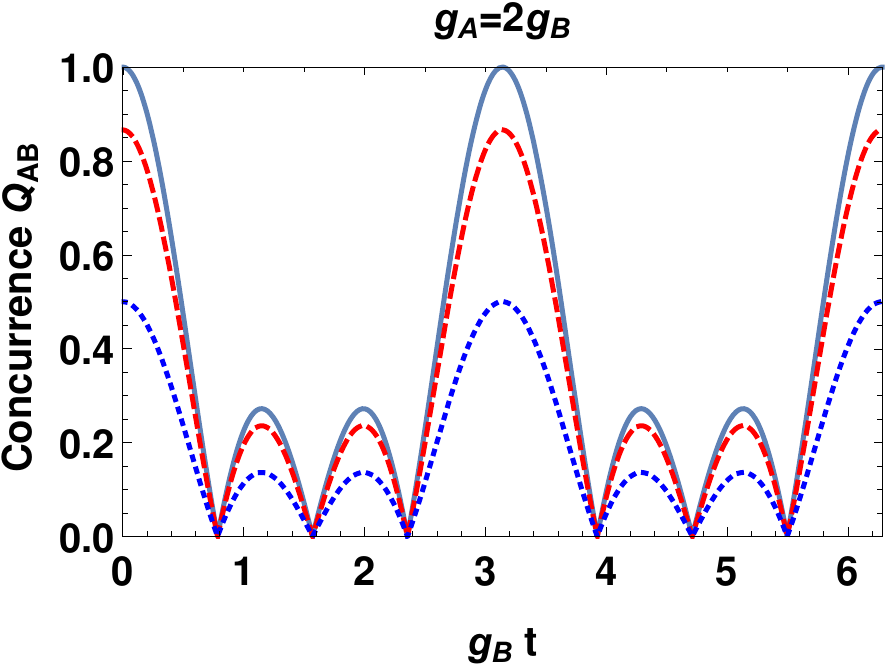}}
\subfigure[]{\includegraphics[width=7cm]{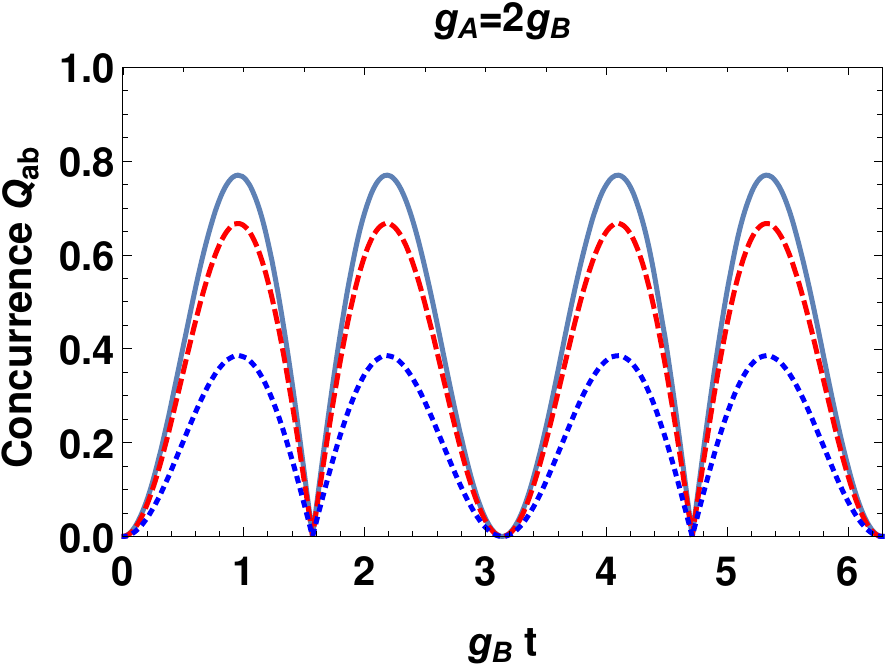}}
\caption{The figures (a)(c)(e) show the evolution of the function $Q_{AB}\left(
t\right) $ of the concurrence of two atoms with time when the initial state
is $\left\vert \psi _{AB}\left( 0\right) \right\rangle $. and (b)(d)(f) express $Q_{ab}\left(
t\right) $ when the initial state$\left\vert \psi _{ab}\left( 0\right) \right\rangle$. The ratio of the coupling strengths take the value of $\frac{g_{A}}{g_{B}}=1$ in subgraphs (a)(b), $\frac{g_{A}}{g_{B}}=\sqrt{2}$ in subgraphs (c)(d), and $\frac{g_{A}}{g_{B}}=2$ in subgraphs (e)(f). In both subgraphs, the gray-blue solid lines represent the initial-state parameter $\theta =\frac{\pi }{4}$, the red dashed lines represent $\theta =\frac{\pi }{6}$, and the blue dotted lines represent $\theta =\frac{\pi }{12}$.
}
\label{fig2}
\end{figure*}

The entanglement information between the two atoms is contained in the
reduced density matrix $\rho ^{AB}$. It can be obtained
by tracing out the photonic parts of the total pure state in Eq.~(\ref{eq-4}). The
reduced density matrix $\rho ^{AB}$ in the basis $\{\left\vert \uparrow \uparrow \right\rangle ,\left\vert \uparrow \downarrow
\right\rangle ,\left\vert \downarrow \uparrow \right\rangle ,\left\vert
\downarrow \downarrow \right\rangle \}$ is given by
\begin{equation}
\label{eq-6}
\rho ^{AB}=\left(
\begin{array}{cccc}
0 & 0 & 0 & 0 \\
0 & \left\vert x\right\vert ^{2} & xy^{\ast } & 0 \\
0 & yx^{\ast } & \left\vert y\right\vert ^{2} & 0 \\
0 & 0 & 0 & \left\vert z\right\vert ^{2}+\left\vert k\right\vert ^{2}%
\end{array}%
\right).
\end{equation}
which is of X-type. We use the concurrence to measure the entanglement between the two atoms~\cite{Hill97}. It is obtained as
\begin{equation}
\label{eq-8}
\mathbb{C}\left( \rho ^{AB}\right) =2\left\vert x\right\vert \left\vert
y\right\vert.
\end{equation}
where the time-dependent probability amplitudes reads
\begin{eqnarray}
\label{eq-9}
\left\vert x\right\vert  &=&\sqrt{\left\vert x_{0} \right\vert ^{2}\cos
^{2}\left( g_{A}t\right) +\left\vert z_{0} \right\vert ^{2}\sin ^{2}\left(
g_{A}t\right) },  \nonumber \\
\left\vert y\right\vert  &=&\sqrt{\left\vert y_{0} \right\vert ^{2}\cos
^{2}\left( g_{B}t\right) +\left\vert k_{0} \right\vert ^{2}\sin ^{2}\left(
g_{B}t\right) }.  
\end{eqnarray}
As all subsystems are two-state systems in the subspace, the following six kinds of entanglement between two-qubit can both be derived, $\mathbb{C}^{AB}$, $\mathbb{C}^{ab}$, $\mathbb{C}^{Aa}$, $\mathbb{C}
^{Bb}$, $\mathbb{C}^{Ab}$, and $\mathbb{C}^{Ba}$. There are some relations between these concurrence~\cite{Sainz07}. But we confine our
attention to $\mathbb{C}^{AB}$.

\section{\label{Sec:3} The case with initial partial Bell states}

In the case of two zero initial coefficients, two subsystems are initially entangled, while the other two subsystems were separable. In principle, there are six possibilities for two coefficients to be zero. These initial states can be expressed as the superpositions of two subsystems Bell states:$\left\vert \psi _{AB}^{\pm }\right\rangle \thicksim \left\vert \uparrow
\downarrow \right\rangle \pm \left\vert \downarrow \uparrow \right\rangle
,\left\vert \psi _{ab}^{\pm }\right\rangle \thicksim \left\vert
10\right\rangle \pm \left\vert 01\right\rangle ,\left\vert \psi _{Aa}^{\pm
}\right\rangle \left( \left\vert \psi _{Bb}^{\pm }\right\rangle ,\left\vert
\psi _{Ab}^{\pm }\right\rangle ,\left\vert \psi _{Ba}^{\pm }\right\rangle
\right) \thicksim \left\vert \uparrow 0\right\rangle \pm \left\vert
\downarrow 1\right\rangle$, respectively. Here we denote the superposition states within each type as follows:
\begin{eqnarray}
\label{eq-10}
\left\vert \psi _{AB}\right\rangle  &=&\cos \theta \left\vert \uparrow
\downarrow \right\rangle +\sin \theta \left\vert \downarrow \uparrow
\right\rangle ,  \nonumber \\
\left\vert \psi _{ab}\right\rangle  &=&\cos \theta \left\vert
10\right\rangle +\sin \theta \left\vert 01\right\rangle ,  \nonumber \\
\left\vert \psi _{Aa}\right\rangle  &=&\cos \theta \left\vert \uparrow
0\right\rangle +\sin \theta \left\vert \downarrow 1\right\rangle .
\end{eqnarray}
Although six different kinds of bipartite entanglements may arise, we will mainly study the entanglement dynamics of two JC atoms with different initial states. And we find among the six different initial states, two kinds of initial states $\left\vert \psi _{Aa}\right\rangle $ and $\left\vert \psi _{Bb}\right\rangle $ need not be considered.

\subsection{\label{Sec:1} Partially entangled Bell states $\left\vert \psi _{AB}\right\rangle $ or $\left\vert \psi _{ab}\right\rangle $ }

\begin{figure*}[htbp]
\centering
\subfigure[]{\includegraphics[width=7cm]{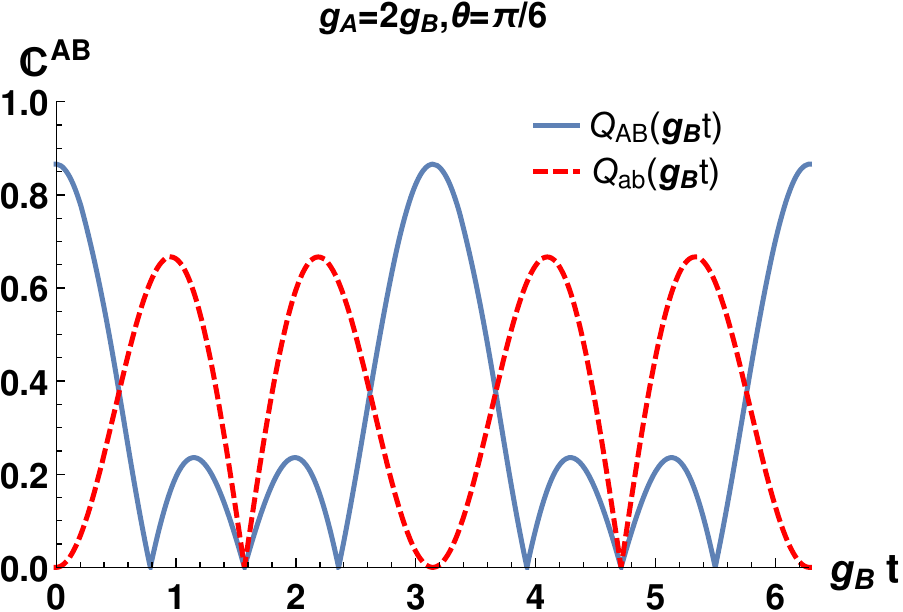}}
\subfigure[]{\includegraphics[width=7cm]{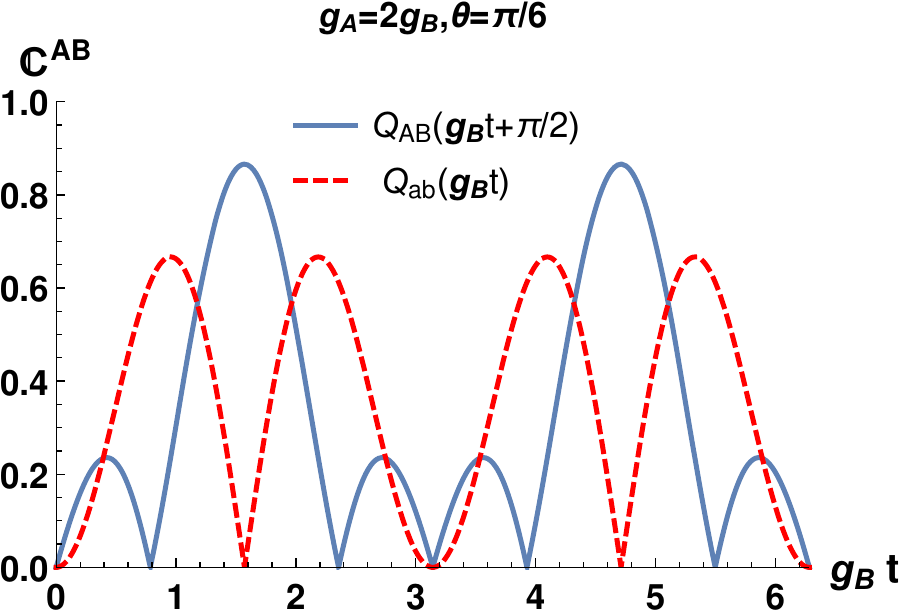}}
\subfigure[]{\includegraphics[width=7cm]{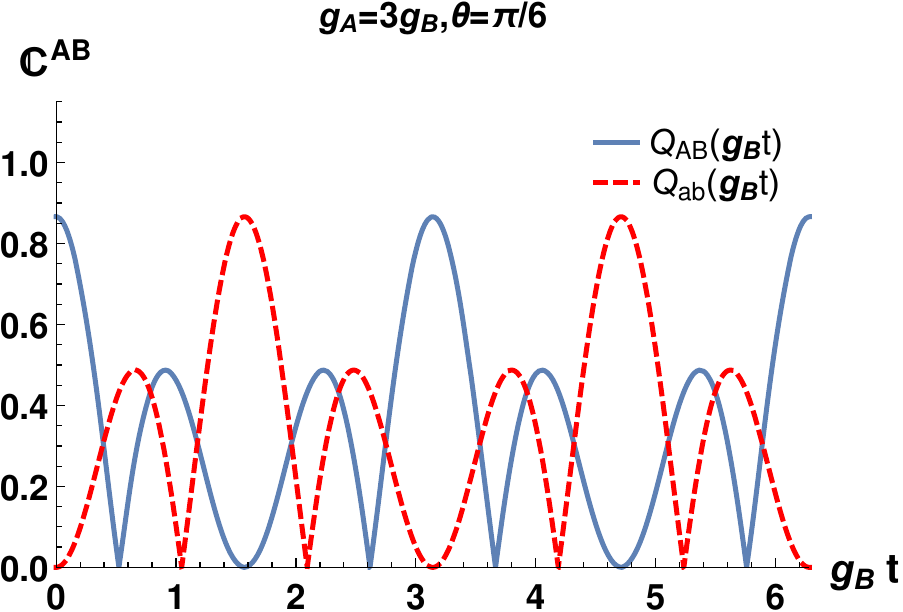}}
\subfigure[]{\includegraphics[width=7cm]{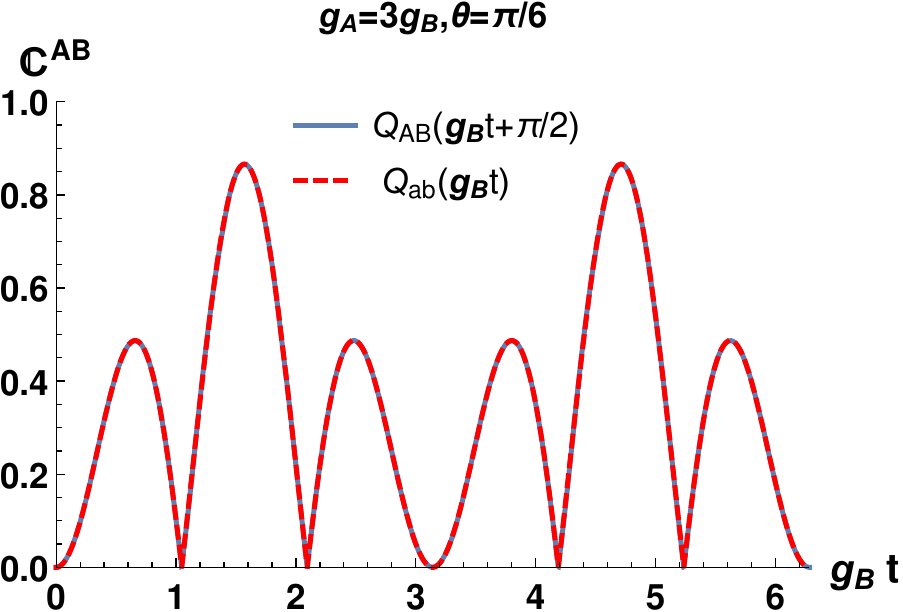}}

\caption{The comparison of atomic concurrence $\mathbb{C}^{AB}$ with two different kinds of initial states. The gray-blue solid lines shows the evolution process of the concurrence for the initial state $\left\vert \psi _{AB}\left( 0\right) \right\rangle $, while the red dashed lines depict the case with the initial state $\left\vert \psi _{ab}\left( 0\right) \right\rangle $. The initial-state parameter $\theta $ in all subgraphs are fixed at $\frac{\pi }{6}$. In subgraphs (a) and (c), the ratio of coupling strengths $\frac{g_{A}}{g_{B}}$ take the value of $ 2, 3$ respectively. Subgraphs (b),(d) are  a comparison between $Q_{ab}\left( gt\right) $ and $Q_{AB}\left( gt\right) $ after shifting the phase to the left by $\frac{\pi }{2}$.
}
\label{fig3}
\end{figure*}

In this subsection, we take the partially entangled Bell states $\left\vert \psi _{AB}\right\rangle $ or $\left\vert \psi _{ab}\right\rangle $ as our initial states. The initial states for the total system reads
\begin{eqnarray}
\left\vert \psi _{AB}\left( 0\right) \right\rangle  &=&\left( \cos \theta
\left\vert \uparrow \downarrow \right\rangle +\sin \theta \left\vert
\downarrow \uparrow \right\rangle \right) _{AB}\otimes \left\vert
00\right\rangle _{ab},  \label{eq-11} \\
\left\vert \psi _{ab}\left( 0\right) \right\rangle  &=&\left( \cos \theta
\left\vert 10\right\rangle +\sin \theta \left\vert 01\right\rangle \right)
_{ab}\otimes \left\vert \downarrow \downarrow \right\rangle _{AB}.
\end{eqnarray}

\begin{figure*}[htbp]
\centering
\subfigure[]{\includegraphics[width=7cm]{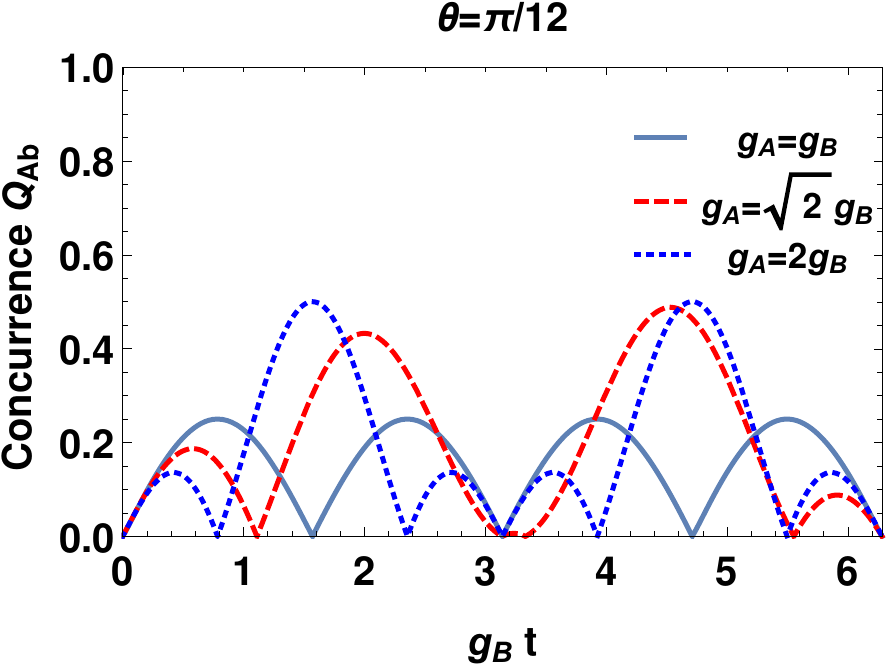}}
\subfigure[]{\includegraphics[width=7cm]{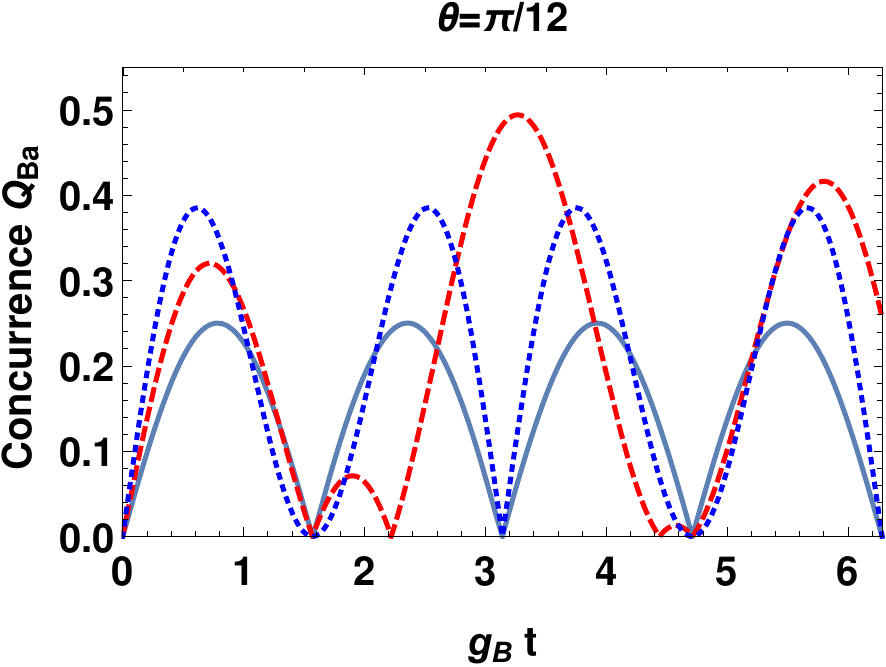}}

\caption{The dynamics of concurrence $\mathbb{C}^{AB}$ for (a) initial state $\left\vert \psi _{Ab}\left( 0\right) \right\rangle$ and (b) $\left\vert \psi _{Ba}\left( 0\right) \right\rangle$. In images (a) and (b) ,the gray-blue solid line, the red dotted line and the blue dot line are respectively used to indicate that the ratio of coupling intensity of the two cavities is 1,$ \sqrt{2}$, 2. The initial-state parameter $\theta $ in all subgraphs are fixed at $\frac{\pi }{12}$. }
\label{fig4}
\end{figure*}

In both cases, the concurrence between atoms is
\begin{equation}
\mathbb{C}^{AB}\left( t\right) =Q_{\alpha \beta }\left( t\right) .
\label{eq-13}
\end{equation}%
with $Q_{\alpha \beta }\left( t\right) $ being the following expression for
the initial state $\left\vert \psi _{\alpha \beta }\left( 0\right)
\right\rangle $:
\begin{eqnarray}
Q_{AB}\left( t\right)  &=&\left\vert \sin \left( 2\theta \right) \cos \left(
g_{A}t\right) \cos \left( g_{B}t\right) \right\vert ,  \label{eq-14} \\
Q_{ab}\left( t\right)  &=&\left\vert \sin \left( 2\theta \right) \sin \left(
g_{A}t\right) \sin \left( g_{B}t\right) \right\vert .
\end{eqnarray}%
It can be clearly seen that the concurrence dynamics between two JC atoms
are determined by the initial-state parameter $\theta $ and the coupling
strengths $g_{A}$, $g_{B}$.

In Fig.~\ref{fig2}, we plot the dynamics of concurrence $\mathbb{C}^{AB}$ with different initial-state parameter $\theta$ and coupling strengths $g_{A}$, $g_{B}$ for initial state $\left\vert \psi _{AB}\left( 0\right) \right\rangle$ or $\left\vert \psi _{ab}\left( 0\right) \right\rangle$. We can see that the zero-concurrence moments depends on the coupling strengths $g_{A}$, $g_{B}$. As we all know, the information in each JC cavity is transferred from the atom to the optical cavity, i.e., $|\uparrow \rangle _{i}$ to $|1\rangle _{i}$, which takes the time of half Rabi periodicity $\frac{\pi }{2g_{i}}$$\left( i=A,B\right) $. When the zero point of entanglement occurs, their must be at least one JC system have completed this transform. What is more, the comparison of all the subgraphs in Fig.~\ref{fig2} shows that the greater the ratio of the coupling strengths, the more the entanglement concurrence fluctuates. These conclusions are consistent with the Yonac's work~\cite{Yonac2006,Yonac07}.

And in Fig.~\ref{fig2}, when the initial states is $\left\vert \psi _{ab}\left( 0\right) \right\rangle$ in Eq.~(\ref{eq-14}). At the initial moment, two JC atoms are in the ground state and are separable. Then with the time, energy is transmitted periodically between the atoms and light fields in the JC models. The $\mathbb{C}^{AB}$ starts from zero and increases to the maximum value, and then collapses and recovers. What is more, the conclusion about the periodicity is the same as that in Fig.~\ref{fig2}. If two Rabi periods are rational, there exists periodic change of atomic entanglement. Otherwise, the periodic phenomenon in the atomic entanglement dynamics vanishes.
\begin{figure*}[htbp]
\centering
\subfigure[]{\includegraphics[width=7cm]{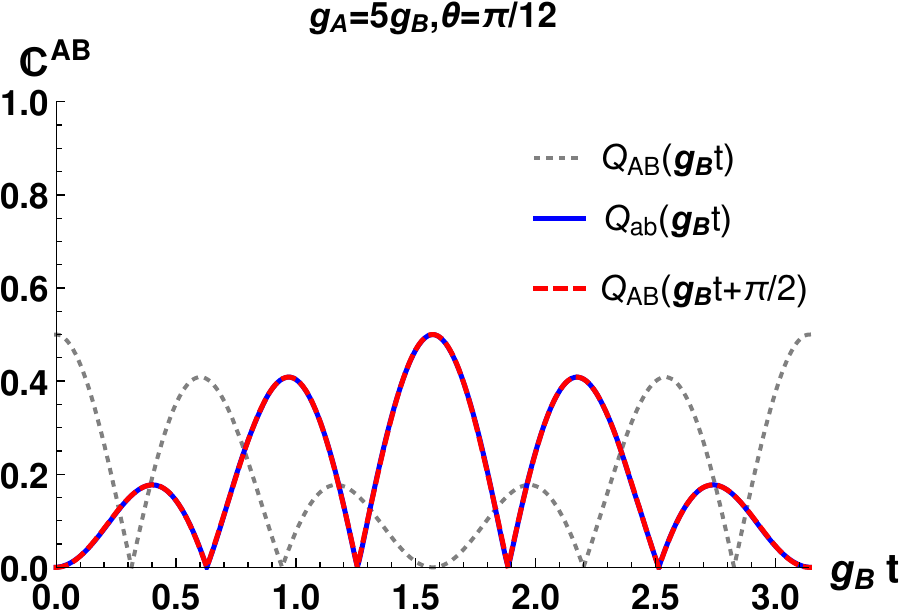}}
\subfigure[]{\includegraphics[width=7cm]{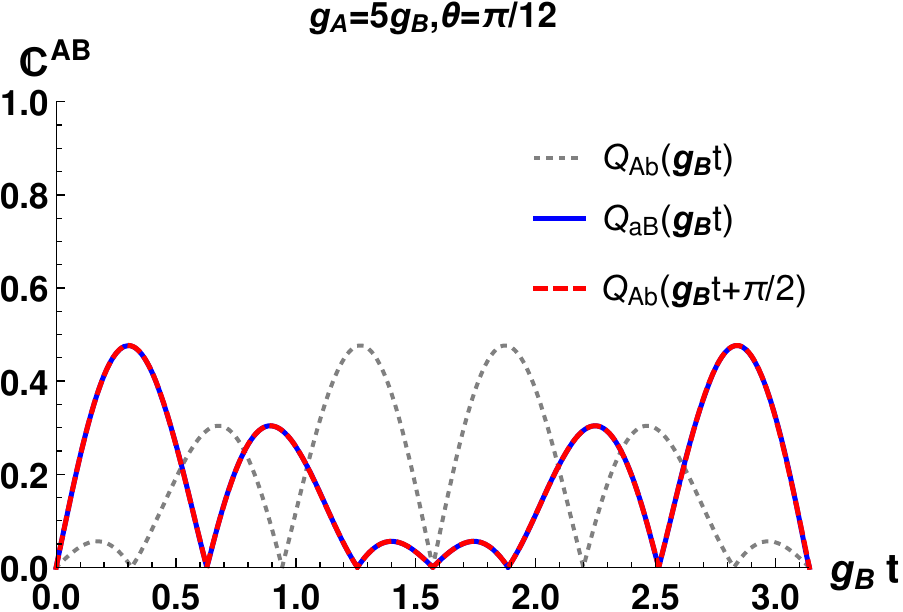}}
\subfigure[]{\includegraphics[width=7cm]{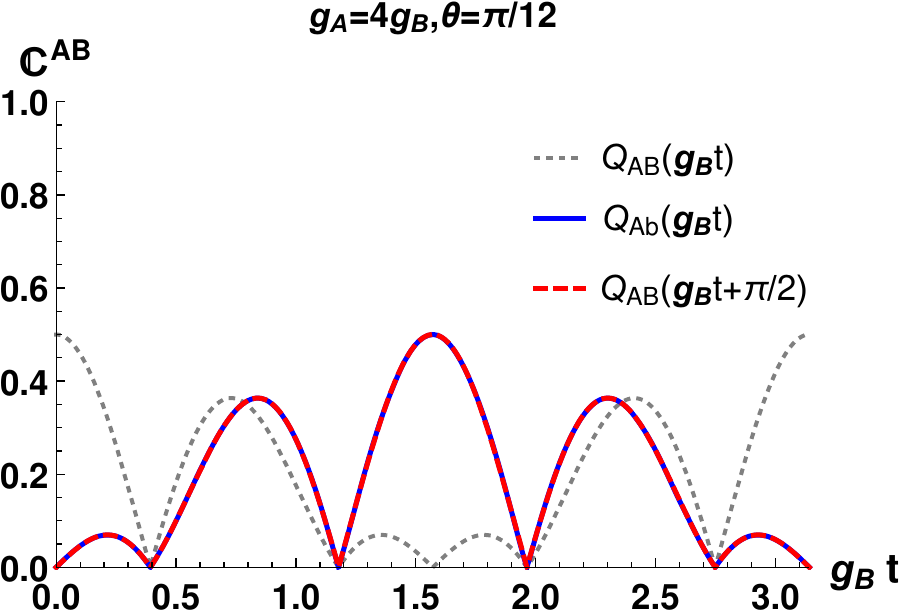}}
\subfigure[]{\includegraphics[width=7cm]{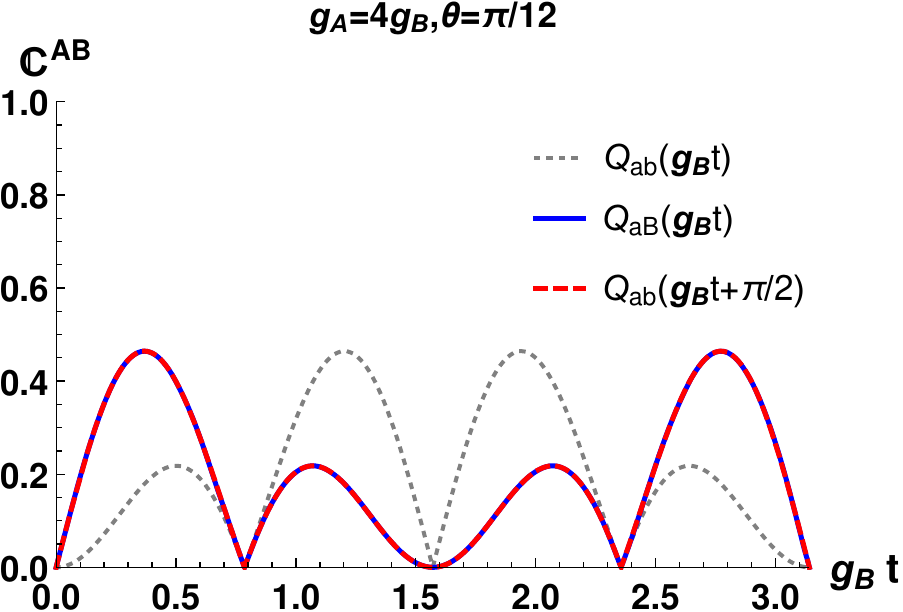}}
\caption{Comparing the entanglement dynamics for four different initial states. The superscript of $Q_{\alpha \beta }$ indicates the case for the initial state $\left\vert \psi _{\alpha\beta}\left( 0\right) \right\rangle$. The ratio of coupling strengths takes the value of $\frac{g_{A}}{g_{B}}=5$ in subgraph (a) and (b), $\frac{g_{A}}{g_{B}}=4$ in subgraphs (c) and (d). In all subgraphs, the initial-state parameter $\theta$ is fixed at $\frac{\pi }{12}$. }
\label{fig5}
\end{figure*}

Now we focus on the periodicity of the concurrence dynamics in this paper. The ratio of the coupling strengths take the value of $\frac{g_{A}}{g_{B}}=1$ in Fig.~\ref{fig2}(a)(b), $\frac{g_{A}}{g_{B}}=2$ in Fig.~\ref{fig2}(e)(f), and $\frac{g_{A}}{g_{B}}=\sqrt{2}$ in Fig.~\ref{fig2}(c)(d). Observing the Fig.~\ref{fig2}(c)(d), the most obvious difference from other subgraphs is that the collapse and recovery of concurrence is no longer periodic. This phenomenon can be explained from the physical perspective. Only under the transformation $\left\vert \uparrow \downarrow 00\right\rangle \rightarrow \left\vert\downarrow \downarrow 10\right\rangle \rightarrow \left\vert \uparrow\downarrow 00\right\rangle $ and $\left\vert \downarrow \uparrow 00\right\rangle \rightarrow \left\vert\downarrow \downarrow 01\right\rangle \rightarrow \left\vert \downarrow\uparrow 00\right\rangle $, the state remains unchanged. This conversion takes the time which is the least common multiple of two Rabi cycles, i.e., $T=k_{A}\frac{\pi }{g_{A}}=k_{B}\frac{\pi }{g_{B}}$ with integers $k_{A}$ and $k_{B}$. This means that the evolutionary periodicity of the $\mathbb{C}^{AB}$ can be found only when the ratio of the two coupling strengths is a rational number $\frac{g_{A}}{g_{B}}=\frac{k_{A}}{k_{B}}$. Otherwise, there is no period in the time evolution of the concurrence.

Besides, the physical mechanism of concurrence dynamics can also be understood from the perspective of energy transfer. For the initial state $\left\vert \psi _{AB}\left( 0\right) \right\rangle $, the energy is distributed in the two atoms at the initial moment, and the entanglement information takes the maximum value $\sin(2\theta)$. Then with the energy transfer from atoms to the light fields, the entanglement between two atoms is destroyed. In a cycle, the number of zero-entanglement times is determined by the ratio of two coupling strengths. When the ratio $n=\frac{g_{A}}{g_{B}}$ is odd, there will be $n$ entangled zeros, while when $n$ is even, there are $n+1$ zeros.

It is natural to compare the entanglement dynamics for two different initial states $\left\vert \psi _{AB}\left( 0\right) \right\rangle $ and $\left\vert \psi _{ab}\left( 0\right) \right\rangle $. As shown in Fig.~\ref{fig3}, when the ratio of coupling strengths is odd, the concurrence with these two initial states differs only $\frac{\pi }{2}$ in phase, as shown in subgraphs (c) and (d). However, the subgraphs (a) and (b) show that the entanglement with these two initial states does not coincide after a simple time translation. This is because the ratio of the evolution periods in the two cavities is even. In this case, the energy of one JC model is already distributed in the light field, but the energy of the other JC model is distributed in the atom. Thus, there is no way to overlap by shifting $\frac{\pi }{2}$ in phase.

\subsection{\label{Sec:2} Partially entangled Bell states $\left\vert \psi _{Ab}\right\rangle $ or $\left\vert \psi _{Ba}\right\rangle $ }

In this subsection, we will further analyze the cases with initial states $\left\vert \psi _{Ab}\right\rangle $ and $\left\vert \psi _{Ba}\right\rangle $. The initial states of the total composite system are
\begin{eqnarray}
\label{eq-16}
\left\vert \psi _{Ab}\left( 0\right) \right\rangle
&=&\left( \cos \theta \left\vert \uparrow 0\right\rangle +\sin \theta
\left\vert \downarrow 1\right\rangle \right) \otimes \left\vert \downarrow
0\right\rangle,\\
\left\vert \psi _{Ba}\left( 0\right) \right\rangle
&=&\left( \cos \theta \left\vert \uparrow 0\right\rangle +\sin \theta
\left\vert \downarrow 1\right\rangle \right) \otimes \left\vert \downarrow
0\right\rangle.
\end{eqnarray}
Similarly to the previous subsection, we also define $Q_{Ab}\left( t\right) $ and $Q_{Ba}\left( t\right) $ as the concurrence of the two atoms with the initial states $\left\vert \psi _{Ab}\right\rangle $ and $\left\vert \psi _{Ba}\right\rangle$, respectively,
\begin{eqnarray}
\label{eq-18}
Q_{Ab}\left( t\right) &=&\left\vert \sin \left( 2\theta \right) \cos \left( g_{A}t\right) \sin
\left( g_{B}t\right) \right\vert,\\
Q_{Ba}\left( t\right)
&=&\left\vert \sin \left( 2\theta \right) \sin \left( g_{A}t\right) \cos
\left( g_{B}t\right) \right\vert.
\end{eqnarray}

Fig.~\ref{fig4} show the entanglement dynamics between atoms for initial states $\left\vert \psi _{Ab}\left( 0\right) \right\rangle $ and $\left\vert \psi _{Ba}\left( 0\right) \right\rangle $. We can see that the periodicity depends on the rationality of the ratio of two coupling strengths. Besides, whether the maximum value can reach $1$ is determined by the initial parameters and the ratio of two coupling strengths.

In Fig.~\ref{fig5}, whether there is time translation between these two kinds of initial states $\left\vert \psi _{Ab}\left( 0\right) \right\rangle $ and $\left\vert \psi _{Ba}\left( 0\right) \right\rangle $ depends on the odd-even number of the coupling strength ratio.  And another interesting conclusion can be obtained by comparing the above four different initial states. When the ratio of two coupling strengths $\frac{g_{A}}{g_{B}}$ is odd, $Q_{AB}\left( gt+\frac{\pi }{2}\right) =Q_{ab}\left( gt\right) $ and $Q_{Ab}\left( gt+\frac{\pi }{2}\right) =Q_{aB}\left( gt\right) $. While when the ratio $\frac{g_{A}}{g_{B}}$ is even, $Q_{AB}\left( gt+\frac{\pi }{2}\right) =Q_{Ab}\left( gt\right) $ and $Q_{ab}\left( gt+\frac{\pi }{2}\right) =Q_{aB}\left( gt\right) $.

\section{\label{Sec:5} CONCLUTION}

We studied the entanglement dynamics of two atoms in the double JC model. The two atoms are coupled to their single-mode optical cavities, and the two JC models are isolated from each other. In the single-excitation subspace, the double JC model can be equivalent to a four-qubit system. In this paper, we use the concurrence to measure the atomic entanglement and consider of initial states is the partial Bell states. We demonstrate that there exist collapses and recovers in the entanglement dynamics. The physical mechanism behind the entanglement dynamics is the periodical information and energy exchange between atoms and light fields. Besides, for the initial Partial Bell states, the evolutionary cycle of the atomic entanglement can be found only if the ratio of two atom-cavity coupling strengths is a rational number. And whether there is time translation between two kinds of initial partial Bell states depends
on the odd-even number of the coupling strength ratio. In summary, our results reveal the dynamic evolution of two-body entanglement in the double JC model in details. And we will further study the entanglement dynamics in multi-excitation space, which can make the light field contain two or more photons to observe the phenomenon of sudden entanglement.

\section*{Acknowledgements}
This work is supported by National Natural Science Foundation of China (Grants No.11935006,and No.11975059).


\begin{thebibliography}{1}

\bibitem{Horodecki2009} R. Horodecki, P. Horodecki, M. Horodecki, and K. Horodecki, Quantum entanglement, Rev. Mod. Phys. 81,865 (2009).
\bibitem{Breuer2002}  H. P. Breuer and F. Petruccione, \emph{The theory of open quantum systems} (Oxford University Press, New York, 2002).
\bibitem{Suter2016}  D. Suter and G.A. Alvarez, Protecting quantum information against environmental noise, Rev. Mod. Phys. 88,041001 (2016).

\bibitem{Walter13} S. Walter, J.C. Budich,  J. Eisert,  and B. Trauzettel, Entanglement of nanoelectromechanical oscillators by Cooper-pair tunneling. Phys. Rev. B 88, 035441 (2013).
\bibitem{Ludwig10} M. Ludwig, K. Hammerer,  and F. Marquardt, Entanglement of mechanical oscillators coupled to a nonequilibrium environment. Phys. Rev. A 82, 012333 (2010).
\bibitem{Joshi12} C. Joshi, J. Larson, M. Jonson, E. Andersson, and P. Ohberg, Entanglement of distant optomechanical systems. Phys. Rev. A. 85, 033805 (2012).
\bibitem{Togan10} E. Togan,  Y. Chu, A.S. Trifonov, L.Jiang, J.Maze, L.Chidress, M.V.G.Dutt, and A.S.SOrensen, Quantum entanglement between an optical photon and a solid-state spin qubit, Nature 466, 730 (2010).
\bibitem{Chathavalappil19}  N. Chathavalappil and S.V.M. Satyanarayana,  Schemes to avoid entanglement sudden death of decohering two qubit system, Eur. Phys. J. D. 73, 36 (2019).

\bibitem{Yu04} Y. Ting and J.H. Eberly, Finite-time disentanglement via spontaneous emission, Phys. Rev. Lett. 93, 140404 (2004).
\bibitem{Yu06} Y. Ting and J.H. Eberly, Quantum Open System Theory: Bipartite Aspects, Phys. Rev. Lett. 97, 140403 (2006).
\bibitem{Yu09} Y. Ting and J.H. Eberly, Sudden Death of Entanglement, Science 323, 598 (2009).
\bibitem{Kimble2007} J. Laurat, K.S. Choi, H. Deng, C.W. Chou, and H.J. Kimble, Heralded Entanglement between Atomic Ensembles: Preparation, Decoherence, and Scaling, Phys. Rev. Lett. 99, 180504 (2007).
\bibitem{Ficek06}  Z. Ficek and R. Tanas, Dark periods and revivals of entanglement in a two-qubit system, Phys. Rev. A 74, 024304 (2006).
\bibitem{Ficek08}  Z. Ficek and R. Tanas, Delayed sudden birth of entanglement, Phys. Rev. A 77, 054301 (2008).
\bibitem{Ou07} Y.C. Ou and H. Fan, Monogamy inequality in terms of negativity for three-qubit states, Phys. Rev. A 75, 062308 (2007).


\bibitem{Qiang18} W.C. Qiang, G.H. Sun, Q. Dong, O. Camacho-Nieto, and S.H. Dong, Concurrence of three Jaynes–Cummings systems, Quantum Inf. Process. 17, 90 (2018).
\bibitem{Jaynes1963} E.T. Jaynes and F.W. Cummings, Comparison of quantum and semiclassical radiation theories with application to the beam maser, Proc. IEEE, 51, 89 (1963).
\bibitem{Eberly1980} J.H. Eberly, N.B. Narozhny, and J.J. Sanchez-Mondragon, Phys. Rev. Lett. 44, 1323 (1980).
\bibitem{Yonac2006}  M. Yonac,  T. Yu, and J.H. Eberly, Sudden death of entanglement of two Jaynes-cummings atoms, J. Phys. B Atomic Mol. Phys. 39, 621 (2006).
\bibitem{Boca04} A. Boca, R. Miller, K. M. Birnbaum, A. D. Boozer, J. McKeever, and H. J. Kimble, Phys. Rev. Lett. 93, 233603 (2004).
\bibitem{Yonac07} M. Yonac, T. Yu, and J.H. Eberly, Pairwise Concurrence Dynamics: A Four-Qubit Model[J], Journal of Physics B Atomic Molecular and Optical Physics, 40, 9 (2007).
\bibitem{Hill97} S. Hill and W.K. Wootters, Entanglement of a Pair of Quantum Bits, Phys. Rev. Lett. 78, 5022 (1997).
\bibitem{Sainz07} I.Sainz, B.G, Entanglement invariant for the double Jaynes-Cummings model[J], Phys. Rev. A. 76(4), 538 (2007).

\end{thebibliography}
\end{document}